\documentstyle[12pt,world_sci]{article}

\newcommand{\ds}{\Delta S=2}
\newcommand{\ve}{\varepsilon}

\newcommand{\als}{\alpha_{s}(s)}
\newcommand{\aps}{\frac{\als}{\pi}}

\newcommand{\smvs}{\vbox{\vskip 8mm}}

\newcommand{\La}{\Lambda_{\overline{MS}}}

\newcommand{\be}{\begin{equation}}
\newcommand{\ee}{\end{equation}}
\newcommand{\ba}{\begin{array}{c}}
\newcommand{\ea}{\end{array}}
\newcommand{\beqn}{\begin{eqnarray}}
\newcommand{\eeqn}{\end{eqnarray}}

\newcommand{\bi}{\begin{itemize}}
\newcommand{\ei}{\end{itemize}}

\newcommand{\rms}{\rm\scriptsize}

\begin{document}

\begin{flushright}
hep-ph/9409222\\
August 1994  \end{flushright}

\title{{\bf QCD corrections to inclusive $\Delta S=1,2$ transitions}}
\author{Matthias JAMIN\thanks{Invited talk presented at the
conference ``QCD'94'', Montpellier, France, July 7 - 13, 1994.
To appear in the proceedings.}\\
{\em Division TH, CERN, 1211 Geneva 23, Switzerland}}

\maketitle
\setlength{\baselineskip}{2.6ex}

\begin{center}
\parbox{13.0cm}
{\begin{center} ABSTRACT \end{center}
{\small \hspace*{0.3cm} The talk summarises a calculation of the two-point
functions for \hbox{$\Delta S=1$} current-current and QCD-penguin operators,
as well as for the $\ds$ operator, at the next-to-leading order.\cite{JP:94}
The size of the gluonic corrections to current-current operators is large,
providing a qualitative understanding of the observed enhancement in
$\Delta I=1/2$ transitions. In the $\Delta S=2$ sector the QCD corrections
are quite moderate ($\approx -20\%$). This work has been done in
collaboration with Antonio Pich.}}
\end{center}


\section{Introduction}
Standard Model calculations of non-leptonic weak decays, conventionally
being performed in the framework of the operator product expansion,
are subject to strong interaction corrections, both for the short-distance
contributions (coefficient functions) as well as for the long-distance
part (matrix elements). Whereas the coefficient functions are evaluated
at a high scale ${\cal O}(M_W)$, and therefore calculable in perturbation
theory, the matrix elements receive contributions from low scales
${\cal O}(\Lambda_{QCD})$ and hence have to be calculated in some
``non-perturbative'' framework.

The situation very much changes if one considers weak decays at the
inclusive level. At the inclusive level, the quantity of interest is
the two-point function of two effective Hamiltonians, which can be
investigated with perturbative QCD methods. In the following, we
shall restrict ourselves to the $\Delta S=1$ Hamiltonian \cite{BJLp:93}
\begin{equation}
\label{eq:hamiltonian}
{\cal H}^{\Delta S=1}_{\mbox{\rms eff}} \; = \;
{G_F\over\sqrt{2}}\,V_{ud}^{\phantom{*}} V_{us}^{*} \sum_{i} C_{i}(\mu^2)
   \, Q_{i} \,,
\end{equation}
obtained through the operator product
expansion, and consider the two-point function
\begin{equation}
\label{eq:correlator}
\Psi^{\Delta S=1}(q^2) \; \equiv \; i \int \! dx \, e^{iqx} \,
\big<0\vert \, T\{\,{\cal H}^{\Delta S=1}_{\mbox{\rms eff}}(x)\,
{\cal H}^{\Delta S=1}_{\mbox{\rms eff}}(0)^\dagger\}\vert0\big> \,.
\end{equation}
The associated spectral function
$\Phi(s)\equiv\frac{1}{\pi}\mbox{\rm Im}\Psi^{\Delta S=1}(s)$
is a quantity with definite physical information.
It describes in an inclusive way how the weak Hamiltonian couples
the vacuum to physical states of a given invariant mass.
General properties like the observed enhancement of $\Delta I=1/2$
transitions can be then rigourously analysed at the inclusive level.
\cite{PDp:91}

\section{Current-current operators}

The two-point functions for the current-current operators are most
conveniently calculated in the diagonal basis in which the two
operators $Q_+$ and $Q_-$ do not mix. The scheme- and
scale-independent spectral functions are then found to be\cite{JP:94}
\begin{eqnarray}
\Phi_{++}(s) & = & \theta(s)\,
\frac{8}{15}\,\frac{s^4}{(4\pi)^6}\,\alpha_s(s)^{-4/9}\,
C_{+}^2(M_W^2) \biggl[\,1-\frac{3649}{1620}\,\aps\,\biggr] \,,
\label{eq:3.31} \\
\smvs
\Phi_{--}(s) & = & \theta(s)\,
\frac{4}{15}\,\frac{s^4}{(4\pi)^6}\,\alpha_s(s)^{8/9\phantom{-}}\,
C_{-}^2(M_W^2) \biggl[\,1+\frac{9139}{810}\,\aps\,\biggr] \,,
\label{eq:3.32}
\end{eqnarray}
where $C_\pm(M_W^2)$ are the coefficient functions corresponding
to the operators $Q_\pm$.\cite{BW:90} All actual calculations have
been performed in two different schemes for $\gamma_5$, an ``naively''
commuting $\gamma_5$ and the algebraically consistent non-anticommuting
$\gamma_5$ according to 't~Hooft and Veltman\cite{HV:72}, and we have
explicitly verified the scale- and scheme-independence of physical
quantities.

Taking $\alpha_s(s)/\pi\approx 0.1$, at the NLO we find a moderate
suppression of $\Phi_{++}$ by roughly 20\%, whereas $\Phi_{--}$
acquires a huge enhancement of the order of 100\%. Because $\Phi_{++}$
solely receives contributions from $\Delta I=3/2$, and $\Phi_{--}$ is
a mixture of both $\Delta I=1/2$ and $\Delta I=3/2$, this pattern of
the radiative corrections entails a strong enhancement of the $\Delta I=1/2$
amplitude. Hence, we are provided with a very promising picture for the
emergence of the $\Delta I=1/2$--rule.

\section{Full result including penguins}
In the process of mixing of operators under QCD corrections, four
additional operators, the so called ``penguin-operators'' are
generated. At the leading order, the resulting effective Hamiltonian
can still be diagonalized, but at the next-to-leading order, this
is no longer possible. So, in this case, the two-point function of
the set of operators is a $6\times6$-matrix.

Following the notation of ref.~[2], the coefficient functions for
$\Delta S=1$ weak processes can be decomposed as $C(s) = z(s) + \tau\,y(s)$,
where $\tau\equiv - \left(V_{td}^{\phantom{*}} V_{ts}^*\right)/
\left(V_{ud}^{\phantom{*}} V_{us}^*\right)$.  The coefficient function $z(s)$
governs the real part of the effective Hamiltonian, and $y(s)$ parametrises
the imaginary part and governs, e.g., the measure for direct CP-violation
in the $K$-system, $\ve'/\ve$. We thus can form two spectral-functions
$\Phi_{z,\,y}(s)$ corresponding to $z$ and $y$ respectively.

In the region $s=1-10\,\mbox{\rm GeV}^2$, relevant for example in QCD
sum rule calculations of the $K$-system, and for a central value
$\La=300\,\mbox{\rm MeV}$, the radiative QCD correction to $\Phi_z$ ranges
approximately between 40\% and 120\%, whereas in the case of $\Phi_y$ we
find a correction of the order of 100\%--240\%.\cite{JP:94} Because the
two-point function is constructed as the square of the effective Hamiltonian,
the actual corrections to ${\cal H}^{\Delta S=1}_{\mbox{\rms eff}}$ are only
about half the corrections to the 2--point function. Therefore, the
perturbative QCD correction to the real part of the effective Hamiltonian
turns out to be 20\%--60\%, and for the imaginary part 50\%--120\%.

\section{The $\ds$ operator}

For the case of $\ds$ transitions, things are somewhat simpler because
there is only one operator. In 4 dimensions this operator is given by
\begin{equation}
Q_{\ds} \; \equiv \; \left( \bar s d \right)_{\rm V-A}
                     \left( \bar s d \right)_{\rm V-A} \,.
\label{eq:6.7}
\end{equation}

Calculating the spectral function $\Phi_{\ds}(s)$ for $Q_{\ds}$,
we obtain
\begin{equation}
\Phi_{\ds}(s) \; = \; \theta(s)\,
\frac{32}{15}\,\frac{s^4}{(4\pi)^6}\,
\alpha_s(s)^{-4/9} \biggl[\,1-\frac{3649}{1620}\,\aps\,\biggr] \,.
\label{eq:6.11}
\end{equation}
Because both $Q_{+}$ and $Q_{\ds}$ have the same chiral representation,
as expected, apart from a global factor, their spectral functions
agree. We observe that the NLO QCD-correction is negative and of the
order of 20\%, for $\als/\pi\approx 0.1$.

\section{Conclusion}

Our work improves and completes the two-point function evaluation
of ref.~[3] with two major additions:
the recently calculated NLO corrections to the Wilson-coefficient functions
have been taken into account and, in addition, we have incorporated
previously missing contributions from evanescent operators.
The final results are then renormalization scheme- and scale-independent
at the NLO, and, therefore, constitute the first complete calculation
of weak non-leptonic observables at the NLO, without any hadronic
ambiguity.

The structure of the radiative corrections to two-point functions of
$\Delta S=1$ and $\ds$ operators also allows for a deeper understanding,
why a description of non-leptonic weak decays in terms of diquarks
was so successful as far as the $\Delta I=1/2$--rule is concerned.
\cite{NSp:91} The explicit calculation shows that quark-quark correlations
give the dominant contribution to the QCD corrections, and through working
with effective diquarks, these contributions can be summed up to all orders.

A full QCD calculation has been possible because of the inclusive
character of the defined two-point functions.
Although only qualitative conclusions can be directly
extracted from these results, they are certainly important since they
rigourously point to the QCD origin of the infamous $\Delta I=1/2$--rule,
and, moreover, provide valuable information on the relative importance
of the different operators, which can be very helpful to attempt
more pragmatic calculations.

\bigskip \noindent
{\bf Acknowledgements}
The author would like to thank S. Narison for the invitation to
the conference and A. Pich for most enjoyable collaboration.


\end{document}